\documentclass{emulateapj}
\usepackage{natbib}
\newcommand{\chan}{{\it Chandra}}

\newcommand{\rf}{${r}_{\rm{F}}$}

\newcommand{\kms}{$\,\rm{km\,s^{-1}}$}
\newcommand{\msolar}{$\rm{M}_{\odot}$}
\shorttitle{Substructure in Abell~3667}
\shortauthors{Owers et al.}
\begin{document}
\title{Substructure in the cold front cluster Abell~3667}
\author{Matt S. Owers\altaffilmark{1}, Warrick J. Couch\altaffilmark{2}, Paul E.J. Nulsen\altaffilmark{3}}
\altaffiltext{1}{School of Physics, University of New South Wales, Sydney, NSW 2052, Australia; mowers@phys.unsw.edu.au}
\altaffiltext{2}{Center for Astrophysics and Supercomputing, Swinburne University of Technology, Hawthorn, VIC 3122, Australia}
\altaffiltext{3}{Harvard Smithsonian Center for Astrophysics, 60 Garden Street, Cambridge, MA 02138, USA}

\begin{abstract}
We present evidence for the existence of significant substructure in the cold front cluster Abell~3667 based on 
multi-object spectroscopy taken with the 3.9m Anglo Australian Telescope. This paper is the second in a series analyzing 
the relationship between cold fronts observed in \chan\, X-ray images and merger activity observed at optical wavelengths.
We have obtained 910 galaxy redshifts in the field of Abell~3667 out to 3.5\,Mpc, of which 550 are confirmed cluster 
members, more than doubling the number of spectroscopically confirmed members previously available and probing some 
3\,mag down the luminosity function. From this sample, we derive a cluster redshift of $z=0.0553\pm0.0002$ and velocity 
dispersion of $1056\pm38$\kms\, and use a number of statistical tests to search for substructure. We find significant 
evidence for substructure in the spatial distribution of member galaxies and also in the localized velocity distributions 
and, in spite of this evidence, find the global velocity distribution does not deviate significantly from a Gaussian. 
Using combined spatial and velocity information, we find the cluster can be separated into two major 
structures, with roughly equal velocity dispersions, but offset in peculiar velocity from each other by $\sim500$\kms, 
and a number of minor substructures. We propose two scenarios which explain the radio and X-ray observations. Our data 
show the cold front is directly related to cluster merger activity, and also highlights the extent of optical data required
to unambiguously detect the presence of substructure. 
\end{abstract}


\section{Introduction}

At the pinnacle of the structure formation hierarchy, major mergers of clusters involve the largest, 
most massive virialized objects at the present epoch. In a major merger, up to $10^{64}$ erg of gravitational potential 
energy is dispersed among the cluster constituents, with around $15\%$ going into the gas through shock heating, 
dissipation of turbulence and adiabatic compression of the intracluster medium (ICM). The majority of the energy goes 
into inducing peculiar motions of the galaxies and dark matter, and a small fraction into cosmic ray acceleration and 
magnetic field amplification. Despite some evidence pointing to rapid evolution of galaxies caused by a major merger 
\citep{caldwell1997}, the implications of mergers for galaxy evolution are not well understood, partly 
because of the difficulty in identifying major mergers, and also in identifying galaxies belonging to substructure. To 
disentangle the physics behind each of these merger driven processes in the ICM, and also to robustly identify 
substructure within clusters, a multi-wavelength approach is required.

To this end, we have selected a sample of clusters exhibiting robust examples of cold fronts from the \chan\, archive 
\citep{Owers_thesis}, and this is the second in a series of papers exploring the sample by employing a 
multi-wavelength analysis. In the first paper \citep[][hereafter Paper I]{owers2008}, we present a combined X-ray and 
optical analysis of Abell~1201, finding multiple substructures through combination of redshift and spatial information 
for 321 cluster members and confirming that the cold fronts observed are a direct result of ongoing merger activity. Here,
we turn our attention to Abell~3667 which is an integral component of this ongoing study, given it was one of the first 
cold fronts detected by \chan\, \citep{vikhlinin2001b} and helped to define the cold front phenomenon. The largest 
spectroscopic study to date was carried out by \citet{johnstonhollitt2008} and based on 231 spectroscopically confirmed 
cluster members. They find no evidence for dynamical activity in the velocity distribution, nor any evidence of localized 
deviations from the global velocity structure and conclude that in order to rule out the existence of substructure, there 
is a need for a larger sample of cluster member redshifts. The current study uses more extensive optical data for 
Abell~3667, with the aim of unambiguously detecting dynamical substructure in Abell~3667 and relating it to the cold front.

At X-ray wavelengths Abell~3667 is luminous 
\citep[$L_X(0.4-2.4 \rm{\,keV})=5.1\times 10^{44}\rm{erg\, s}^{-1}$;][]{ebeling1996}, hot 
\citep[${kT} \sim 6$\,keV;][]{knopp1996} and displays several features indicative of a major merger. The ROSAT 
X-ray observations of \citet{knopp1996} suggested the cluster was not in dynamical equilibrium, with an overall elliptical 
appearance, central structure and significant X-ray emission associated with the second D galaxy to the north-west. 
Perhaps the most interesting feature in the X-ray images is the surface brightness discontinuity to the south-east of the 
cluster center initially interpreted as a merger shock by \citet{markevitch1999}. However, the increase in both spectral 
and spatial resolution and sensitivity with the \chan\, X-ray observatory provided the necessary data 
to conclude that the discontinuity is a cold front - a contact discontinuity at the interface of dense, cold gas cloud 
moving at near sonic speed through the hotter, less dense ICM \citep{vikhlinin2001b}. The striking sharpness of the front
suggests transport processes across the front must be suppressed, and the stability of the front has led to interesting 
insights into the growth of hydrodynamic instabilities due to the motion of the gas cloud within the ICM 
\citep{vikhlinin2001a,churazov2004}.

Abell~3667 exhibits more interesting signatures of an ongoing merger at radio wavelengths, and distinguishes itself as 
one of only a few clusters with two diffuse radio relics positioned symmetrically on opposing sides of the cluster 
periphery with their major axis perpendicular to the direction of cluster elongation \citep{rottgering1997}. 
Radio relics are steep spectrum, filamentary, polarized, synchrotron sources and \citet{rottgering1997} hypothesized that 
turbulence in the cluster outskirts associated with a cluster merger was responsible for re-accelerating electrons to the 
relativistic speeds needed to produce the observed synchrotron emission. However, \citet{roettiger1999} were able to 
produce the observed properties of the radio relics by simulating a cluster merger which produced shock fronts that inject
the energy required to accelerate electrons to ultra-relativistic speeds required for the synchrotron radiation 
\citep[also see][and references therein]{hoeft2007}. More indirect evidence for a merger
comes in the form of a head--tail galaxy located $\sim 635$\,kpc to the north-west of the central galaxy 
\citep{rottgering1997}, which according to \citet{bliton1998} are preferentially found in dynamically 
active clusters.

In the optical, Abell~3667 manifests itself as a richness class 2, Bautz-Morgan I-II intermediate type cluster 
\citep{abell1989} at redshift $z=0.055$. \citet{girardi1998} measure a virial mass of $1.2\times 10^{15}h^{-1}$\,\msolar 
and velocity dispersion $971$\kms\, for 154 members within $2.2\,h^{-1}\,$Mpc. Earlier work by \citet{sodre1992} and 
\citet{proust1988} measure higher dispersions (1100-1400\kms) but with smaller samples. Despite the clear evidence of 
merger activity at X-ray and radio wavelengths, the existence of substructure detected at 
optical wavelengths remains ambiguous. Multimodality has been observed in projected galaxy density maps 
\citep{sodre1992,proust1988}, although this was not detected at high significance in the density maps of 
spectroscopically confirmed cluster members in \citet{johnstonhollitt2008}. Based on the DEDICA substructure detection 
algorithm, \citet{ramella2007} detect a number of substructures in Abell~3667 using spatial plus color information, whilst 
\citet{joffre2000} use weak lensing to produce a projected mass map which is strongly peaked around the central D galaxy, 
with three other peaks at lower significance - one to the north, one coincident with the second bright D galaxy and one 
to the south-east of the cluster center.

In this paper, we present multi-object spectra (MOS) obtained with AAOmega at the 3.9m Anglo Australian 
Telescope (AAT), more than doubling the number of spectroscopically confirmed members of 
Abell~3667. In \S\ref{datared} we present the photometric catalog, selection criteria for MOS observations and data 
reduction. In \S\ref{membership} we present the method for cluster membership determination. We outline the methods used 
for detection of substructure in \S\ref{substructure} and in \S\ref{discussion} we discuss and summarize our analysis.
Throughout the paper, we assume a standard $\Lambda \rm{CDM}$ cosmology where $H_0=70$\kms, $\Omega_m=0.3$ 
and $\Omega_{\Lambda}=0.7$. For this cosmology and at the redshift of the cluster $1''=1.075$ kpc.

\section{Optical sample selection and observations}\label{datared}

In this section, we present the method for selection of photometric candidates for spectroscopic follow up, details of the 
observations conducted at the 3.9m Anglo-Australian Telescope (AAT) and brief descriptions of data reduction, redshift 
determination, redshift accuracy and spectroscopic completeness.

\subsection{Parent Photometric Catalog}\label{parent.cat}

The parent photometric catalog was obtained from the Supercosmos Sky Survey (SSS)
server\footnote{see: http://www-wfau.roe.ac.uk/sss/} using the object catalog extraction function. Objects classified 
as galaxies (SSS class=1) within a radius $R=53'$ (3.4\,Mpc) of the central dominant cluster 
galaxy at RA=$20^{\rm h} 12^{\rm m} 27.38^{\rm s}$, DEC=$-56^{\circ}$ 49$\arcmin$ 35.7$\arcsec$ were included. We further 
filtered the catalog such that for $R \leq 2$\,Mpc, only objects with magnitudes $r_F \leq 19.0$ were included, whereas 
for $2 < R \leq 3.4$\,Mpc a brighter magnitude limit of $r_F \leq 18.0$ was adopted. These selection criteria limited the
catalog to a manageable size, given the available telescope time, while retaining the objects most likely to be cluster
members. This resulted in a total of 2,163 galaxies being included in our parent catalog. 

\begin{figure}
{\includegraphics[angle=-90,width=.48\textwidth]{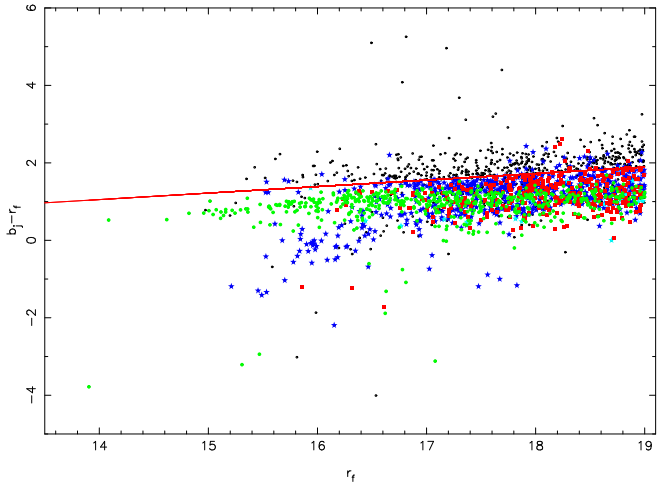}}
\caption{Color-magnitude (CM) diagram for objects in the field of Abell~3667. The {\it green points} represent 
spectroscopically confirmed cluster members (see \S\ref{membership}), the {\it dark blue stars} are foreground stars, the 
{\it light blue stars} are foreground galaxies, and the  {\it red squares} are background galaxies. The {\it black points}
are objects without spectra. The {\it solid line} shows the red envelope we adopted for separating objects which are 
redward of the cluster's red CM sequence and hence are likely non-members,  from those that are on or blueward of the 
red sequence and hence have a high likelihood of being members.}
\label{br.vs.r}
\end{figure}

To maximize the efficiency of our spectroscopic observations in terms of targeting cluster members, we ranked each galaxy
in the parent catalog on the basis of its position in the color-magnitude (CM) plane (see Figure \ref{br.vs.r}), as 
well as its projected distance from the cluster center. The position of galaxies relative to Abell~3667's red 
color-magnitude sequence was used as a means to weight our target selection, with the red sequence being identified 
using 160 spectroscopically confirmed members obtained from the NED catalog. These, and further spectroscopically 
confirmed members as determined in \S~\ref{membership}, are represented by the {\it green} 
points in Figure~\ref{br.vs.r}, and at magnitudes fainter than $r_F\sim 16.5$ are seen to delineate a red sequence which 
exhibits the usual negative slope in the CM diagram. At $r_F<16.5$, this red sequence persists, but exhibits a positive 
slope, which we can only assume is due to some systematic problem in the photometry at these brighter magnitudes. This 
behavior, however, is not problematic for pre-selecting targets, since all we want to do is use the red sequence to 
define a likely upper red envelope, below which most cluster members reside (galaxies above are almost certainly more 
distant non-members). The {\it solid} straight line shown in Figure \ref{br.vs.r} (for which 
$b_J-r_F = 0.168\times r_F-1.771$) represents the red envelope that was adopted. Its position and slope were set to best 
include all galaxies on and bluewards of the red sequence, allowing for the systematic change in its slope and the 
increasing photometric scatter at fainter magnitudes. Galaxies meeting these criteria at $R \le 500$\,kpc from the cluster
center were given the highest rank. For more remote galaxies, the rank was decreased for each additional 500\,kpc step in 
distance from the cluster center. Galaxies lying lying above the red envelope were all given lower rank than those below, 
as well as being ranked in a similar manner by their cluster-centric distances.

For completeness, we also plot in Figure \ref{br.vs.r} objects which have been shown in \S~\ref{membership} to be 
foreground stars (shown as the {\it dark blue stars}), foreground galaxies ({\it light blue stars}), and background 
galaxies ({\it red squares}). The small {\it black points} indicate objects in the field without spectra. 

\subsection{AAOmega Observations}

Substantial new spectroscopy of Abell~3667 was obtained with the Two Degree Field(2dF)/AAOmega multi-object spectrograph 
facility on the 3.9\,m Anglo-Australian Telescope (AAT) during the nights of 2007 July 14-18. AAOmega is a bench-mounted 
dual-beam spectrograph which is fed by 400 fibers which are robotically placed within the AAT's two degree field of view 
at prime-focus \citep{saunders2004, smith2004, sharp2006}. All observations were taken using the medium resolution 
($R\simeq1300$) 580V (blue arm) and 385R (red arm) gratings. In combination with the 2\,arcsec diameter fibers, they 
deliver a spectral resolution at the detector (a 2k$\times$4k $15\mu m$ pixel E2V CCD in each arm) of $3.6$\AA\ in the 
blue and $5.5$\AA\ in the red. The wavelength range covered by the two arms is $3700-8800$\AA.

The large number of galaxy targets and their highly clustered distribution on the sky, together with 30\,arcsec 
restriction on the minimum separation between fibers, meant that six different plate configurations were required to 
obtain a high level of spectroscopic completeness and adequate coverage in the central regions of Abell~3667 where the
cold fronts are located. The fiber allocations for each of these configurations were done automatically using the AAOmega 
specific {\it CONFIGURE} program\footnote{See: http://www.aao.gov.au/AAO/2df/aaomega/aaomega\_software.html}. This program
uses a ``simulated annealing'' algorithm \citep{miszalski2006} to ensure the optimal allocation of fibers in terms of 
maximizing the number of objects observed. In doing so, it uses the rankings/weights assigned to targets 
(\S~\ref{parent.cat}) to determine what priority they have in being allocated a fiber. 

The use of the simulated annealing algorithm required the input catalog be reduced to no more than $\sim 800$ objects 
for fiber allocation. To achieve this, the catalog was first split into the 1452 higher priority objects, on or blueward 
of the of the red sequence, and the 711 lower priority ones, redward of the red sequence (\S~\ref{parent.cat}). The higher
priority group were further subdivided into 830 objects with $r_F \leq 18$ and the 622 with $18<r_F\leq 19$.
Two fiber configurations were required to observe the high priority $r_F \leq 18$ 
catalog. The small number of objects in the $r_F\leq 18$ catalog that were not observed in these two configurations 
were added to the $18<r_F\leq 19$ high priority catalog. Finally we constructed a catalog made up of those objects not
already allocated, plus previously allocated objects with unreliable redshifts (see below) and a small number of low 
priority objects. Hence in total, six configurations were observed, the majority of which targeted objects lying on or 
blue-ward of the Abell~3667's red sequence. Observing conditions were generally cloud-free, but the seeing was rather 
poor, being in the range 2.4-4.0\,arsec (FWHM). However, given the extended nature of our galaxy targets due to their 
relatively low redshifts, mediocre seeing did not have any serious impact on our program. The dates, exposure times, 
magnitudes and conditions for the observations are summarized in Table~\ref{obslog}.

\begin{deluxetable*}{ccccc}
\tabletypesize{\scriptsize}
\tablecolumns{5}
\tablewidth{0pc}
\tablecaption{Summary of the AAT/AAOmega observations for Abell~3667.\label{obslog}}
\tablehead{
\colhead{Date} &\colhead{Object Priority} &\colhead{Magnitude} & \colhead{Frames} & \colhead{Seeing}}
\startdata
 2007 Jul 14 &High&\rf\,$<18.0$&$4 \times 1800\rm s $ &2.8-2.9\arcsec\\
  2007 Jul 16 &High&\rf\,$<18.0$&$3 \times 1800\rm s $ &2.8-3.5\arcsec\\
  2007 Jul 16 &High&\rf\,$<19.0$&$4 \times 1800\rm s $ &2.4-3.0\arcsec\\
  2007 Jul 16,18 &High&\rf\,$<19.0$&$2 \times 1800 + 3 \times 1200\rm s $ &3.5\arcsec\\
  2007 Jul 17,18 &High+Low&\rf\,$<19.0$&$3 \times 900 + 3 \times 1800\rm s $ &2.8-4\arcsec\\
  2007 Jul 18 &High+Low&\rf\,$<19.0$&$3 \times 1200\rm s $ &3.5\arcsec\\
\enddata
\end{deluxetable*}

For sky subtraction, $>40$ fibers per configuration were allocated to blank sky regions, selected using the 
``{\it Generate sky grid}'' command in the {\it CONFIGURE} software and visually inspected to ensure they were free 
of objects. Approximately 35 known ``fringing'' fibers were excluded from the configuration 
process. Prior to each fiber configuration being observed, tungsten and arc lamp exposures were taken for flat-field and 
wavelength calibration, respectively. The data were fully reduced at the telescope using the AAO 
{\sf 2dfdr}\footnote{see: http://www.aao.gov.au/AAO/2df/aaomega/aaomega.html} pipeline reduction 
system. This system extracts and produces fully flat-fielded, wavelength-calibrated and sky-subtracted (but not 
flux-calibrated) spectra for each exposure frame, co-adds the spectra in each exposure set, and then splices together the 
blue and red-arm spectra. 

Redshifts were identified and measured from the reduced spectra using the {\sf runz} code written by Will Sutherland for 
the Two Degree Field Galaxy Redshift Survey \citep[2dFGRS;][]{colless2001}. This program utilizes the cross-correlation 
method of \citet{tonry1979}. Each spectrum was visually inspected and given a redshift quality classification, $Q$, 
ranging from 1-6, as described in Paper I. A $Q$ value of 3, 4, or 5 indicates a reliable redshift was 
obtained (with a larger value indicating a higher quality spectrum), whereas Q values less than 3 indicate either an 
unreliable redshift or no redshift was obtained. A value of $Q=6$ was used in cases where the redshift indicated the 
object to be a star. We obtained 1714 spectra during the run, of which 1570 yielded reliable ($Q>2$) redshift 
measurements containing 910 extragalactic objects and 660 stellar objects ($Q=6$), indicating the unreliability of the SSS 
star/galaxy classifications. 

\subsection{Redshift Measurement Uncertainties}

\begin{figure}
{\includegraphics[angle=-90,width=.48\textwidth]{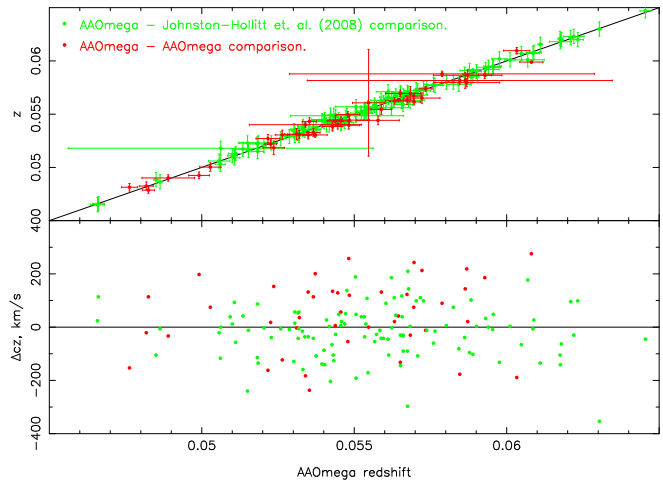}}
\caption{{\it Top panel} Shows a comparison of redshift measurements for multiply observed objects where the 
{\it green} points show AAOmega-\citet{johnstonhollitt2008} redshift comparison and {\it red} points show the comparison
of the AAOmega spectra with double observations. The {\it solid black} line shows the one-to-one relationship. The {\it 
lower panel} shows the redshift difference as a function of redshift where the color coding is as for the upper panel.}
\label{zcomp}
\end{figure}

Of the objects observed twice, 50 had two reliable ($Q>2$) redshift measurements and so in these cases we were able to 
compare redshifts to assess the robustness of our measurements and determine their uncertainty. After excluding one 
obvious outlier where the redshift difference was $\Delta cz=12732$km/s, we found 
$\overline{\Delta cz}= 50\pm20$\,\kms\ for the difference between the first and second groups of measurements. We also 
have 117 high quality measurements
in common with the redshift catalog of \citet{johnstonhollitt2008} and we found $\overline{\Delta cz}= -18\pm9$\kms, 
again after excluding one outlier with $\Delta cz=16134$\,\kms. These redshift comparisons are encapsulated in 
Figure~\ref{zcomp}, noting that we only consider objects in the redshift range 0.045 -- 0.065 (i.e. that 
appropriate to Abell~3667). The upper panel of Figure~\ref{zcomp} shows that the redshift comparisons scatter about a 
one-to-one relationship ({\it solid line} with slope unity), indicating there are no redshift-dependent systematic 
effects. The redshift differences are shown in the lower panel. We derive an RMS redshift difference of 151\kms\ for 
the AAOmega double observations, which implies a redshift uncertainty of 107\kms\, for the sample, comparable to the
RMS of the individual redshift uncertainties which is 112\kms. An external check is given by the RMS of the differences 
between our AAOmega and \citet{johnstonhollitt2008}'s redshifts, which gives an uncertainty of 
64\kms, much lower than the quadrature sum of 148\kms\, for the uncertainty above and the 
uncertainty of the \citep{johnstonhollitt2008} sample ($\sim 100$\kms). We take the precision of our redshift 
measurements to be the more conservative value of 107\kms.

In cases where a galaxy had multiple redshift measurements, the following policy was adopted in determining which value 
would be used for our analysis. If both the measurements were made with AAOmega, we adopted the redshift with the higher 
$Q$ value. If the $Q$ values were equal, then we took the redshift with the smallest measurement uncertainty (as provided by 
the {\sf runz} program). If the second redshift measurement came from \citet{johnstonhollitt2008}'s catalog, we used our 
AAOmega redshift (for consistency). Only 25 extra redshifts were added to our catalog from the \citet{johnstonhollitt2008}
catalog, taking our total number of redshifts to 1595.

\subsection{Redshift Completeness}

\begin{figure*}
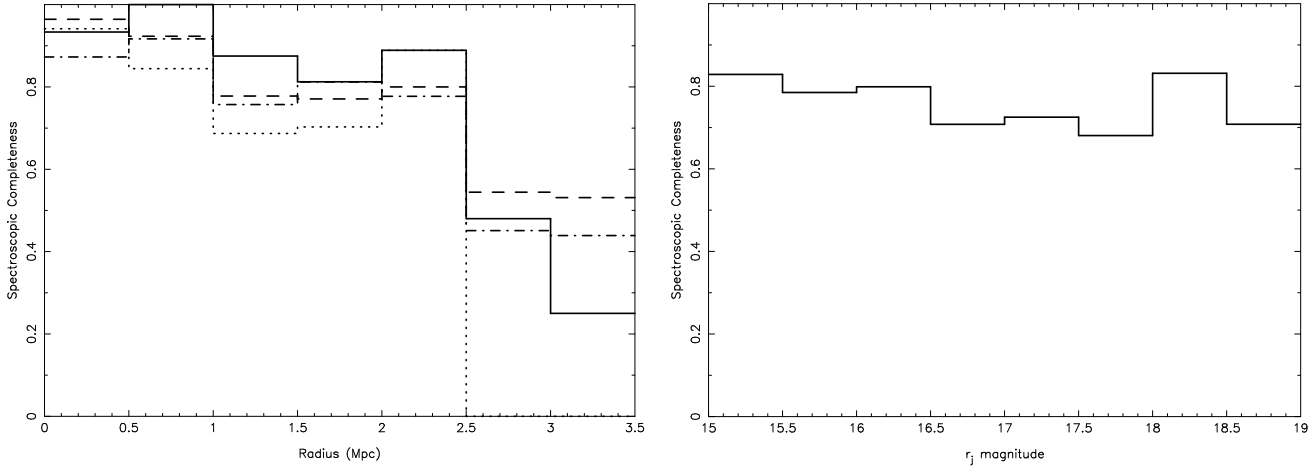

\begin{tabular}{cc}
{\includegraphics[angle=-90,width=0.47\textwidth]{f3a.eps}}
\hspace{1.5 mm}
{\includegraphics[angle=-90,width=0.47\textwidth]{f3b.eps}}
\end{tabular}
\caption{{\it Left:} Spectroscopic completeness as a function of cluster-centric radius, for different magnitude ranges: $r_F<16$ ({\it solid} line), $16\leq r_F<17$ ({\it dashed} line), $17\leq r_F<18$ ({\it dot-dashed} line), and $18\leq r_F<19$ ({\it dotted} line). {\it Right:} Spectroscopic completeness as a function of \rf\, magnitude.}
\label{compl}
\end{figure*}

The completeness of our redshift measurements was determined by calculating the fraction of galaxies in the parent 
photometric catalog for which reliable redshifts were obtained. In the left panel of Figure~\ref{compl} we plot the 
completeness as a function of cluster-centric radius for the magnitude ranges $r_F<16$, $16\leq r_F<17$, $17\leq r_F<18$ 
and $18\leq r_F<19$. It shows that we have achieved excellent completeness within a cluster-centric radius of 2.5\,Mpc. 
We also plot the completeness as a function of $r_F$ magnitude in right-hand panel of Figure \ref{compl} which 
demonstrates a high level of completeness across all magnitude bins. The cluster is very well sampled at magnitudes 
brighter than $r_F=19$ for all radii less than 2.5\,Mpc. If we assume the cluster $M^*_{r_F}=-20.84$ \citep{eke2004} and 
neglect K-correction and galactic extinction terms, we probe $\sim 3$\,mags down the luminosity function.

\section{Cluster member selection}\label{membership}

\begin{figure}
{\includegraphics[angle=-90,width=.48\textwidth]{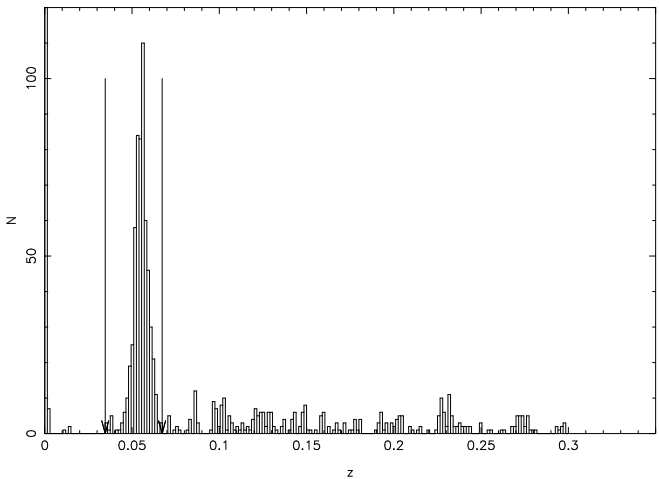}}
\caption{Redshift distribution for all robust measurements in the Abell~3667 field. Abell~3667 clearly stands out against 
the projected background and foreground objects. The arrows indicate the separation of the cluster based on gaps in the 
velocity distribution of $cz=1100$\kms\, (see text). The bin at $z=0$ contains some 660 stellar objects.}
\label{allhist}
\end{figure}

The major driver of this study is the detection of substructure using a sample of spectroscopically confirmed cluster 
members. To achieve this, it is imperative that the cluster member catalog be free of line of sight interlopers which
may produce false substructure. Thus, a robust selection procedure is required, and we
have proceeded in much the same manner as described in Paper I. Briefly, initial rejection of 
foreground and background objects was performed by sorting objects in redshift space, determining the velocity gap 
between neighboring galaxies (here velocity is defined as $cz$) and identifying the cluster 
\citep[via the method of][]{depropris2002} as the peak in redshift space that is separated from foreground and background 
galaxies by a gap greater than some specified value. Here we use 1100\,\kms, which is the velocity dispersion measured by 
\citet{johnstonhollitt2008}. The redshift distribution of galaxies within the Abell~3667 field is shown in 
Figure~\ref{allhist}, where Abell~3667 stands out clearly from foreground and background objects.

\begin{figure}
{\includegraphics[angle=-90,width=.48\textwidth]{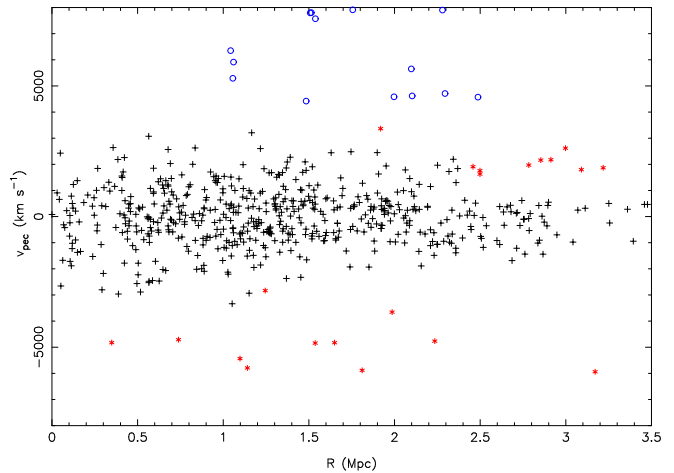}}
\caption{Refinement of cluster membership using the shifting gapper method (see text for details). The {\it black crosses}
are cluster members, {\it red asterisks} indicate objects that are rejected as interlopers and the {\it blue circles} 
indicate objects rejected prior to the shift gapper refinement.}
\label{gapper}
\end{figure}

The membership was further refined using the ``shifting gapper'' method of \citet{fadda1996} which is described in detail
in Paper I. Briefly, the data are binned radially such that each bin contains 50 or more objects and interlopers are 
rejected in each radial bin based on gaps in the peculiar velocity distribution. Figure \ref{gapper} shows the results 
of applying the shifting gapper method, where it is clear that a number of foreground galaxies are rejected with this 
refined procedure, while a number of background galaxies very close in redshift space at radii greater than 2.5\,Mpc are 
also rejected.

\begin{figure}
{\includegraphics[angle=-90,width=.48\textwidth]{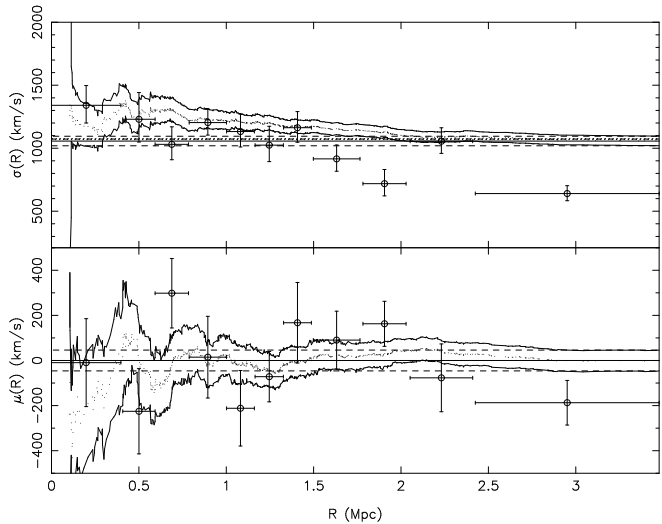}}
\caption{{\it Top panel} shows the integral ({\it points} and 
{\it solid} $1\sigma$ confidence limits) and differential ({\it circles} with $1\sigma$  error bars) $\sigma(R)$ 
measurements. Also plotted are the values measured considering all members ({\it solid line} with {\it dashed} $1\sigma$ 
limits), and the value obtained by assuming the same specific energy in the gas and galaxies and using 
the gas temperature to determine the expected velocity dispersion ({\it dot-dashed} line with {\it dotted} $90\%$ 
confidence limits). {\it Bottom panel} shows the corresponding $\mu(R)$ measurements.}
\label{vparmprofs}
\end{figure}

Our final cluster sample contains 550 members within a cluster-centric radius of $\sim 3.5$\,Mpc, more than doubling the 
number of spectroscopically confirmed cluster members previously known. Using this expanded sample, we apply the robust 
biweight scale and locations estimators described in \citet{beers1990} to determine a cluster velocity dispersion, 
$\sigma_v=1056\pm38$\kms, and mean a redshift, $z_{cos}=0.0553\pm0.0002$ where the quoted errors are $1\sigma$ values 
and were determined using the jackknife re-sampling technique. Presented in Figure~\ref{vparmprofs} are the integral and differential profiles of the location ($\mu(R)$) and scale ($\sigma(R)$) 
estimators. The differential profiles are determined using the radial bins from the shifting gapper analysis. The 
integral profiles are obtained through sorting the members by cluster-centric radius and measuring $\mu(R)$ and 
$\sigma(R)$ within the projected radius of each galaxy, beginning with the tenth galaxy from the center. The integrated 
$\sigma(R)$ becomes relatively constant outside $R=2$\,Mpc, indicating that the cluster velocity dispersion measured at 
this radius is free from the effects of velocity anisotropy, and is representative of the cluster's gravitational 
potential. The differential $\sigma(R)$ is a declining function of radius, and is significantly lower in the outskirts. 
The differential $\mu(R)$ measurements are noisy, but are generally consistent, within the errors, with the value 
measured for the whole cluster.

\section{Substructure Detection}\label{substructure}

In this section, we search for substructure using the velocity distribution, the distribution of galaxies on the sky and 
a combination of spatial plus velocity information. These various statistical methods have differing sensitivities to the
variety of substructure that can occur in clusters \citep{pinkney1996,girardi2002}. Using a range of tests improves our
chances of detecting any substructure that is present.

\subsection{Quantifying the Gaussianity of the Velocity Distribution}

Dynamically relaxed clusters are expected to have line-of-sight velocity distributions which are very close to Gaussian. 
Departures from Gaussianity in cluster velocity distributions can signify a number of different physical phenomena, eg., 
highly radial orbits, interloper contamination, circular orbits and, most commonly, substructure within the cluster. It 
must be noted, however, that non-detection of departures from Gaussianity is not necessarily evidence of a relaxed 
system \citep[][See also Paper I]{pinkney1996}.

Our initial test for departures from Gaussianity involves the use of the Kolmogorov-Smirnov (K-S) statistic 
\citep{press1992} which measures the largest absolute departure between the cumulative probability distribution of a 
data sample and a theoretical distribution describing the data (here, a Gaussian) and determines the significance of the 
difference by calculating the likelihood that the data are drawn from a parent distribution which is Gaussian, taking 
into account the sample size. The K-S test reveals no evidence for non-Gaussianity at the $50\%$ level.

The disadvantages of the K-S test are its lack of sensitivity in the tails of a distribution (it is most 
sensitive in the vicinity of the median), its spurious results in the presence of interlopers and also its failure to 
quantify the way in which a distribution may differ from Gaussianity, which is clearly important in determining the 
physical driver of the departure from Gaussianity. Thus, we used the method first applied to galaxy cluster velocity 
distributions by \citet{zabludoff1993} where the distribution is approximated by a series of three Gauss-Hermite 
functions: the coefficient of the zeroth order term, $h_0$, multiplies a best-fitting Gaussian, the third order term, 
coefficient $h_3$, describes asymmetric deviations from Gaussianity (similar to skewness) and the fourth order term, 
coefficient $h_4$, describes symmetric deviations (similar to kurtosis). It has the advantage of being insensitive to 
outliers in the tails of the distribution (compared to the Gaussian skewness and kurtosis measurements). For Abell~3667, 
the third and fourth order terms had values of $h_3=0.009$ and $h_4=0.047$. 
\begin{figure}
{\includegraphics[angle=-90,width=.48\textwidth]{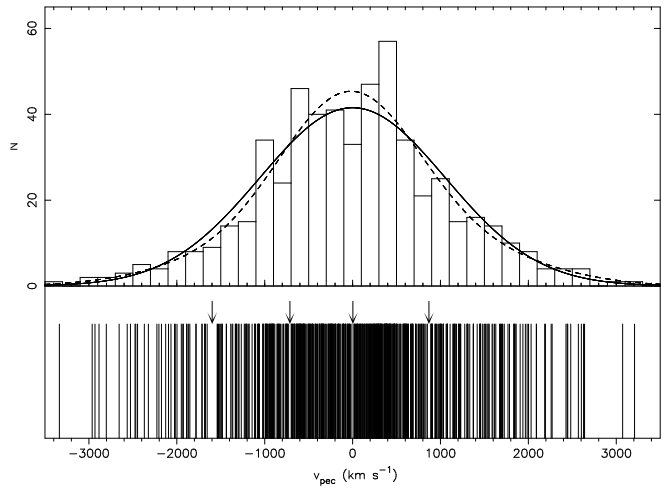}}
\caption{{\it Top panel:} peculiar velocity histogram with bins of width  200\kms. The {\it solid} curve is a Gaussian 
with a mean of zero and standard deviation of 1056\kms\, and the {\it dashed} curve shows the Gauss-Hermite 
reconstruction of the distribution. {\it Bottom panel:} a strip density plot where the arrows indicate the positions of 
weighted gaps with normalized values greater than 2.5 (see text).}
\label{strip.histo}
\end{figure}

To determine the significance of the results, we generated 10,000 Monte-Carlo realizations of Gaussians with $N=550$ and 
mean and dispersion equal to the best fitting value derived for our data, and determined the number of times our 
observed values occurred in the simulations. Values of $|h_3|\ge0.009$ occur in $77\%$ of the realizations, so we conclude 
the observed $h_3$ term is not significantly different from zero. Values of $|h_4|\ge0.047$ occur only $10\%$ of the time, 
thus we conclude the observed $h_4$ term is mildly significant. The positive value of the $h_4$ term means that the 
velocity distribution has slightly longer tails and is more peaked than the best fitting Gaussian, suggestive of either 
radial orbits or substructure. In Figure \ref{strip.histo} we plot the binned velocity distribution, with both a 
Gaussian (with a mean of zero and a standard deviation of $1056$\kms) and the Gauss-Hermite reconstruction curves 
overlaid. 
\begin{figure*}
{\includegraphics[angle=-90,width=\textwidth]{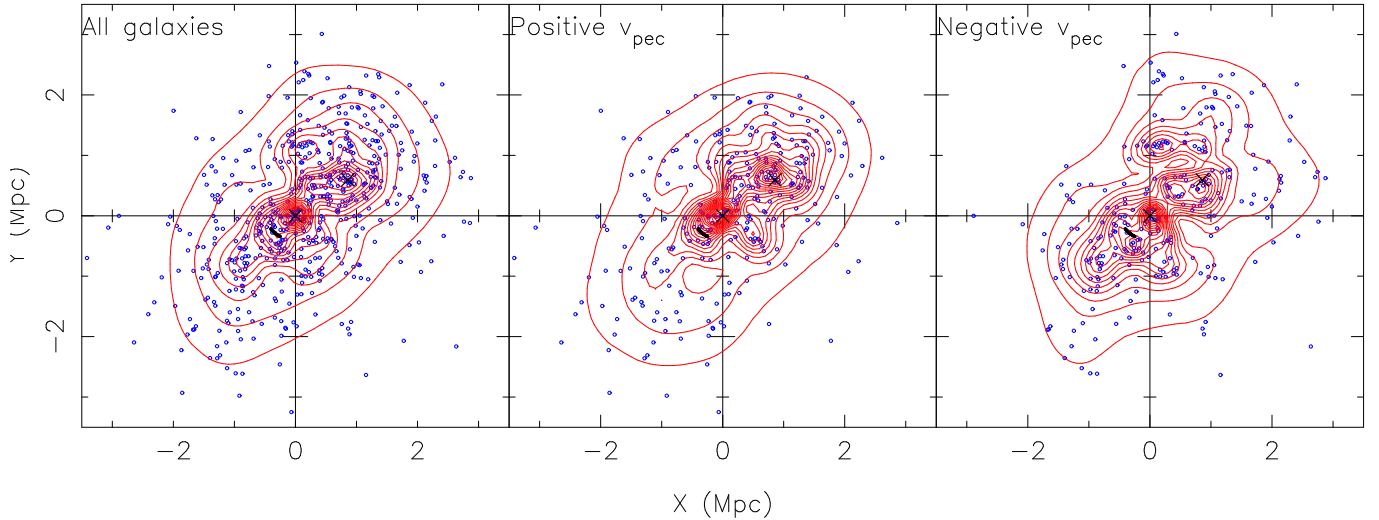}}
\caption{The {\it left-hand} most panel shows the spatial distribution for the entire cluster member population with 
smoothed contours (see text) drawn at 10\,gals\,Mpc$^{-2}$ intervals. The positive and negative velocity maps are shown 
in the {\it middle} and {\it right-hand} most panels, respectively. The contour levels are drawn at intervals of 
5\,gals\,Mpc$^{-2}$. In each panel, black crosses are drawn at the positions of the two dominant cluster galaxies, and 
the black arc in the south-east quadrant marks the position of the cold front.}
\label{wgap.dists}
\end{figure*}

We searched for possible segregation in the velocity distribution using weighted gaps \citep{beers1990}, where the 
velocities are sorted in increasing order and the $i$th velocity gap is given by $g_i=v_{i+1}-v_i$. The weight for the
$i$th gap is $w_i=i(N-i)$ and the weighted gap is defined as $\sqrt {w_i g_i}$. The weighted gaps are normalized through 
dividing by the mid-mean ($MM$) of the ordered weighted gap distribution given by
\begin{equation}
MM = 2/N \sum^{3N/4}_{i=N/4}\sqrt{w_ig_i}
\end{equation}
where $N=550$. Normalized weighted gaps which have values larger than 2.5 occur $1.4\%$ of the time when selecting gaps 
at random from a Gaussian distribution \citep{wainer1978}, thus 8 such gaps are expected for Abell~3667. In fact we find 
4 such gaps, shown as arrows in Figure \ref{strip.histo} where the positions of each peculiar velocity in the sample is 
plotted in a strip density plot. We conclude that the weighted gap analysis does not provide significant evidence for 
segregations in the velocity distribution of Abell~3667.

\subsection{2D galaxy Distribution}\label{galdist}
While the tests for Gaussianity are useful for detecting mergers occurring along our line of sight, where the velocity 
distribution is significantly perturbed, they can be ineffective in the case of a merger occurring near 
the plane of the sky \citep[][and Paper I]{pinkney1996}. In these cases, projected galaxy density maps are useful tools for
searching for evidence of substructure, but are contaminated by background or foreground interlopers projected along the 
line of sight when redshift information is not available. 

Given the presence of significant substructure apparent in the projected galaxy density maps of \citet{proust1988} and 
\citet{sodre1992}, and the non-detection in of significant substructure in the maps of spectroscopically confirmed 
cluster members in \citet{johnstonhollitt2008}, we use our much larger sample to produce a galaxy density map to search 
for signs of substructure. We plot the distribution of all galaxies in the sample, along with the associated surface 
density contours in the left-hand panel in Figure~\ref{wgap.dists}. The contours were generated by smoothing the 
distribution with a variable-width Gaussian, where $\sigma$ varies from $140$\,kpc in the center to $500$\,kpc in the 
outskirts. We find similar results to \citet{proust1988} and \citet{sodre1992}, finding Abell~3667's galaxy surface 
density distribution is clearly bimodal, with a high surface density core at the center and also a high-density region 
offset by $\sim 1$\,Mpc to the north-west which is coincident with the second dominant cluster galaxy. There are two other 
clear peaks in the surface density, one $\sim 1.2$\,Mpc to the south-east and one $\sim 1.1$\,Mpc just west of due north.

As a first order test for correlation between multi-peaked surface density distribution and the velocity distribution,
we split the catalog into positive and negative peculiar velocities and plot the spatial distributions. The two 
segregations are shown in the {\it middle} and {\it right-hand} panels of Figure~\ref{wgap.dists}, along with the 
associated galaxy surface density contours, generated using the same Gaussian smoothing described above. The most notable 
features upon comparing the positive and negative maps are; the south-east and north-east peaks, which are clearly more 
significant in the negative peculiar velocity map and the north-west peak appears in both the positive and negative 
maps, although it is more significant in the positive velocity map and slightly offset in the negative velocity map.

\subsection{3-D tests for sub-structure}\label{3dstructures}

In the previous section, we searched for spatial/velocity correlations by crudely splitting the catalog into positive and
negative velocity samples. In this section, we combine both velocity and spatial information to test for substructure in 
a more refined manner. In general it has been found that tests simultaneously employing both spatial and velocity 
information are most successful at detecting substructures \citep{pinkney1996}. Here, we incorporate the k-statistic, 
hereafter $\kappa_s$, which was first used in the analysis of substructure within Coma by \citet{colless1996}. The $\kappa_s$ 
tests for departures between local and global kinematics by determining the probability that the local velocity 
distribution differs from the global one. 
In brief, the evaluation of $\kappa_s$ is performed as follows: The nearest $n=\sqrt N$ 
neighbors in projected radius from each galaxy in the sample are selected, where $N$ is the number of cluster members. 
The velocity distribution of the $n$ neighbors is compared to the velocity distribution of the remaining $N-n$ galaxies 
using the K-S D-statistic and the likelihood of finding a D-statistic larger than the observed D-statistic is determined. 
Hence, low likelihoods give an indication that the local and global velocity distributions are significantly different. 
The negative log of the likelihoods are summed to give an overall measure of the substructure present within the cluster. 
The significance of the substructure is determined by repeating the above analysis for 10,000 Monte Carlo realizations, 
where any correlation between the spatial and velocity distributions due to substructure within the cluster is erased by 
randomly shuffling the peculiar velocities in the sample, whilst maintaining spatial information. For Abell~3667, we 
measured an overall $\kappa_s$ of 422. In contrast, the $\kappa_s$ distribution produced in our Monte Carlo realizations
is well modeled by a log-normal distribution with a mean of $\mu$($\ln \kappa_s$)= 5.44 and standard deviation 
$\sigma$($\ln \kappa_s$)=0.15, indicating our result is significant at the $3.96\sigma$ level, and a value as 
large as 422 was not obtained in any realization. 

\begin{figure}
{\includegraphics[angle=-90,width=.48\textwidth]{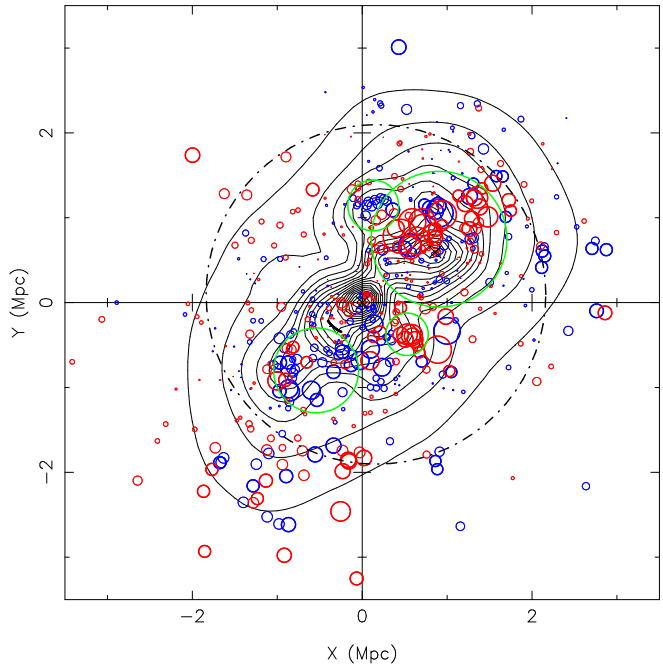}}
\caption{A `bubble' plot of Abell~3667, where the circles represent the size of the k-statistic, which provides a measure 
of the departure of the local velocity distribution from the global one. The circles are colored {\it blue} or {\it red},
depending on whether $v_{pec}$ is negative or positive, respectively. Circles plotted with a thicker line indicate a 
significant departure whereby the k-statistic value occurs only in $5\%$ of 10,000 Monte Carlo realizations. {\it Green} 
circles indicate regions of interest which are used as inputs to the KMM algorithm, and the {\it black dot-dash} circle 
shows the region within which the KMM analysis was done. The black contours are generated as described in \S~\ref{galdist} 
and are plotted at intervals of 10\,gals\,Mpc$^{-2}$ and black crosses are drawn at the positions of the two dominant 
cluster galaxies, and the black arc in the south-east quadrant marks the position of the cold front.}
\label{bubbleplot}
\end{figure}

\citet{pinkney1996} showed that strong radial gradients in the cluster velocity dispersion profiles can produce high 
false positive detection rates for 3D substructure tests. With this in mind, and given that Figure~\ref{vparmprofs} shows 
a significant gradient in the $\sigma(R)$ profiles for Abell~3667, we re-measured $\kappa_s$ within the 2\,Mpc region 
shown in Figure \ref{bubbleplot}. The observed $\kappa_s$ in the smaller region is 312, and a value of this order is 
obtained in only 12 of our 10,000 Monte Carlo realizations which produce a mean of $\mu$($\ln \kappa_s$)=5.27 and 
standard deviation $\sigma$($\ln \kappa_s$)=0.17, indicating the observed $\kappa_s$ is significant at the $3.13\sigma$ 
level, again indicating the existence of substructure within Abell~3667 at high significance. 

The most effective way to illustrate the substructure within the cluster is to present `bubble' plots, with circles whose
radii proportional to the negative log of the measured K-S likelihood are plotted at each cluster member position. 
This is shown in Figure~\ref{bubbleplot}, where regions of clustered large circles indicate the existence of substructure.
Circles in this plot are color-coded, red for galaxies with positive $v_{pec}$, and blue for galaxies with negative 
$v_{pec}$. Also, galaxies having K-S likelihoods which occur in only $5\%$ of the simulated cases are emboldened. The 
circles are plotted on top of the galaxy surface density contours, generated as described in \S\ref{galdist}.

There are a number of regions where the presence of groupings of significantly large `bubbles' indicate localized velocity
structure, confirming our measure of significant substructure being present within Abell~3667. These localized regions of 
velocity structure are generally coincident with surface density enhancements, and, most significantly in terms of the 
cold front, we observe a galaxy density enhancement $700$\,kpc to the south-east of the front which is coincident with a 
clustering of large bubbles, indicating the presence of a subgroup. Interestingly, the significant enhancement in galaxy 
density to the north-west, coincident with the 2nd ranked cluster galaxy, shows little to no deviation in its local 
velocity distribution, despite being coincident with a peak in the projected mass maps of \citet{joffre2000}. 
There is, however, localized velocity structure surrounding the substructure which may be associated.

\begin{figure*}
{\includegraphics[angle=-90,width=.9\textwidth]{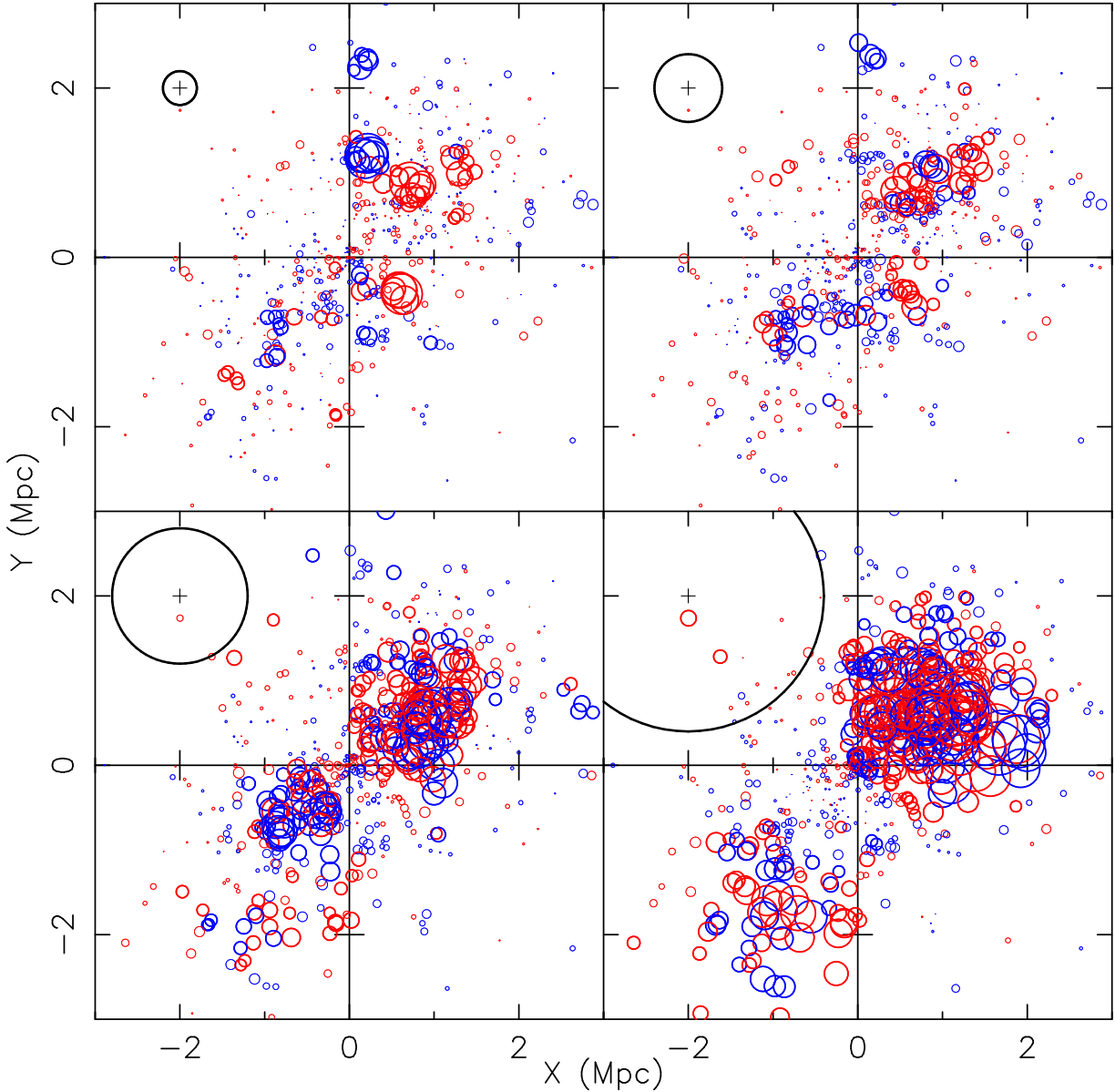}}
\caption{Multi-scale bubble plots where the circle size and color is defined as in Figure~\ref{bubbleplot}, but
rather than using the $n=\sqrt N$ nearest neighbors to define the local velocity distribution, we use the near neighbors
within a fixed projected aperture of $R_{NN}$=200, 400, 800 and 1600\,kpc shown in the ({\it upper left}), 
({\it upper right}), ({\it lower left}) and ({\it lower right}) panels, respectively. The {\it bold black} circle in the 
upper left of each panel has radius $R_{NN}$. }
\label{multiscale}
\end{figure*}

It is possible that the non-detection of localized velocity structure around the region of enhanced galaxy 
density to the north-west is due to the difference in the scale probed by the test compared to the physcial
scale of the substructure. Since the local velocity distribution is defined using a fixed number of near neighbors, 
rather than the near neighbors within some fixed projected radius, it is difficult to determine exactly which scales are 
being probed at each galaxy position in Figure~\ref{bubbleplot}. All that is known is that where the galaxy density is 
high, smaller scales are probed and, conversely, where the galaxy density is low, larger scales are probed. To remedy 
this, and also to address the fact that substructures can have a range of different physical scales, we use a 
multi-scale analysis. This involves remeasuring $\kappa_s$ in a similar fashion to that described above, but rather than 
using a fixed number of $n$ nearest neighbors to define the local velocity distributions, we use the near neighbors 
within a fixed projected radius, $R_{NN}$, of the galaxy of interest. The analysis is repeated for near neighbors within 
$R_{NN}=200, 400, 800$ and 1600\,kpc and the resulting bubble plots are shown in Figure~\ref{multiscale}. For these radii,
we measure $\kappa_s$ values of 342\,($4.15\sigma$), 405\,($3.71\sigma$), 762\,($4.62\sigma$) and 1114\,($4.04\sigma$), 
respectively, where the bracketed values are the significances, measured as above from 10,000 Monte Carlo realizations. 
Across the four scales presented in Figure~\ref{multiscale}, all of the substructures
seen in Figure~\ref{bubbleplot} are reproduced and, most significantly, on scales 800 and 1600\,kpc the north-west 
substructure appears as a region containing significant local velocity substructure.

Our concern regarding the $\sigma(R)$ gradient affecting the $\kappa_s$ measurement seems warranted, given the 
detection of significant velocity structure $>2$\,Mpc to the south-east in Figure~\ref{bubbleplot} and the lower panels of
Figure~\ref{multiscale}. This structure is not offset significantly in peculiar velocity, however its dispersion is
much lower than the total cluster value and therefore it is unlikely to be a bound infalling substructure. Rather it is 
likely to be an artifact of the declining $\sigma(R)$ profile.

Having identified significant substructure within Abell~3667, we now use the KMM (Kaye's Mixture Modeling) algorithm 
\citep{ashman1994} to partition the different substructures. The algorithm fits a user-specified number of Gaussians to a 
data set and determines the improvement of the fit over that of a single Gaussian. Here we utilize the combination of 
spatial and velocity information and fit 3-D Gaussians. Inspection of Figure \ref{bubbleplot} reveals four regions of 
interest (outlined by green circles) where there is localized velocity structure and/or there is structure in the galaxy 
surface density. These regions serve as an initial estimate of the number of partitions into which the KMM algorithm 
should divide the cluster, and we restrict our analysis to the region containing these structures defined by the black 
dot-dash region in Figure \ref{bubbleplot}. Within the regions of interest, obvious velocity outlyers are rejected, and the
mean and standard deviation are measured for the spatial distributions, while the median and median absolute deviation
\citep{beers1990} are measured for the velocity distribution. The measurements serve as initial inputs to the algorithm
and, along with the outputs from the KMM algorithm, are shown in Table \ref{kmmtable}, where the final 
$\overline{v}$ and $\sigma_v$ values are determined using the median and gapper estimators in cases where $N_G < 15$, 
and the biweight location and scale estimators in cases where $N_G \geq 15$ members, respectively. 

\begin{deluxetable*}{cccccccc}
\tabletypesize{\scriptsize}
\tablecolumns{8}
\tablewidth{0pc}
\tablecaption{Input and output parameters for the KMM algorithm. The final column gives the estimates for the correct 
allocation rate for each KMM partition.\label{kmmtable}}
\tablehead{
\colhead{}&\multicolumn{3}{c}{Initial Inputs}& \multicolumn{4}{c}{KMM Outputs}\\
\cline{1-8}\\
\colhead{Group} &  \colhead{($\overline{x},\overline{y}, \overline{v}$)} &\colhead{($\sigma_{x}, \sigma_{y}, \sigma_{v}$)} 
&\colhead{$N_{gal}$}&\colhead{$(\overline{x},\overline{y}, \overline{v})$} &\colhead{($\sigma_{x}, \sigma_{y}, \sigma_{v}$)}
& \colhead{$N_{gal}$}&\colhead{Rate ($\%$)}}
\startdata
KMM1 &(130, 1145, -1388)&(112, 66, 353 ) & 10 &  (153,  1169, -1483)&(83, 58, 395)& 9 & 100\\
KMM2 &(905, 747, 408)&(368, 370, 829)& 110 &(879, 824, 422)&(530, 560, 1039)& 164 &89\\
KMM3 &(528, -373, 1375) & (91, 89, 704) & 9&(563, -415, 1457)&(64, 44, 426)& 7 & 100\\
KMM4 &(-542, -798, -523) &(292, 238, 489)& 36&(-761, -659, -638)&(412, 326, 209)& 27 & 73\\
KMM5 & (0, 0, 55)&(827, 853, 1020)& 284 &( -144, -173, -103)&(676, 739, 1073)& 242 & 87\\
\enddata
\tablecomments{The units of $\overline{x},\overline{y},\sigma_x$ and $\sigma_y$ are kpc, and the units of $\overline{v}$ 
and $\sigma_v$ are \kms. For $\overline{v}$ and $\sigma_v$, the median and gapper estimators are used when $N_G < 15$ and
where $N_G \geq 15$, the biweight location and scale estimators are used.}
\end{deluxetable*}

\begin{figure}
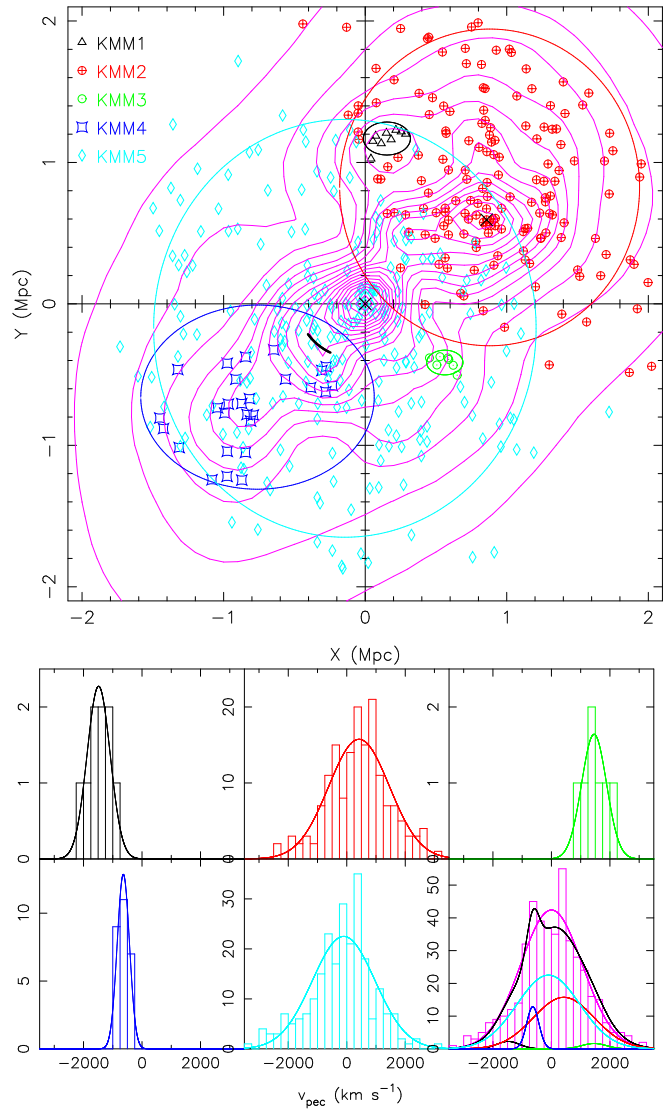

  \begin{center}
      {\includegraphics[angle=-90,width=0.48\textwidth]{f11a.eps}}\\
      {\includegraphics[angle=-90,width=0.47\textwidth]{f11b.eps}}
    \caption{{\it Upper panel:} The spatial distribution of the KMM partitions. The ellipses are the $2\sigma$ boundaries for 
      the Gaussians fitted to the spatial distributions. Galaxy contours with the same spacings as in Figure~\ref{bubbleplot} 
      are shown in {\it pink}, and the color key for the positions is given at upper left. Black crosses are drawn at the 
positions of the two dominant cluster galaxies, and the black arc in the south-east quadrant marks the position of the 
cold front. {\it Bottom panel:} the velocity 
      histograms corresponding to the KMM partitions, color-coded to match the upper panel. At the bottom right we show the 
      velocity distribution for all galaxies included in the KMM analysis, along with the break-down of the Gaussian components 
      making up the total velocity distribution. We also overplot the best fitting single Gaussian ({\it pink curve}) for 
      comparison.}
    \label{kmmalloc}
  \end{center}
\end{figure}

In Figure \ref{kmmalloc}, we show the spatial distribution of the partitions, along with the associated velocity 
histograms for each partition. The cluster is essentially divided into north-west and south-east partitions, with the 
south-east one associated with the dominant cluster galaxy, and the north-west one associated with the 
second ranked galaxy coincident with the second major overdensity in the galaxy surface brightness maps. The velocity 
dispersions of these components are very similar ($1039\pm66$\kms\, for KMM2 and $1073\pm64$\kms\, for KMM5), but their 
velocities are slightly offset, with KMM2 having $\overline v = 422\pm82$\kms, similar to the peculiar velocity of the of 
the second ranked galaxy which is at 432\kms, and KMM5 has $\overline v = -103\pm71$\kms. There are two other minor 
structures (KMM1 and KMM3) which appear to be small bound groups, and a low velocity dispersion structure to the 
south-east (KMM4), which is associated with an overdensity in the galaxy surface density maps.

\section{Discussion and Conclusions}\label{discussion}

Abell~3667 shows three signatures of a recent merger. There is significant substructure in both the spatial and localized 
velocity distributions of the galaxies. It has a pronounced cold front and other structure in its X-ray emitting gas 
\citep{knopp1996,markevitch1999,vikhlinin2001a} and it contains two radio relics \citep{rottgering1997}. Here we consider 
how these structures were created.

\subsection{Merger Scenario}\label{mergescen}

The multiwavelength data at hand may be reconciled with a simple two-cluster merger. The structure to the north-west is
clearly significant in the 2D galaxy density maps and is coincident with the second dominant cluster galaxy. The structure
is also coincident with a significant mass concentration in the weak-lensing maps of \citet{joffre2000}. There is only a 
small offset in peculiar velocity of $\sim 422$\kms, and very little difference in the velocity dispersion between the 
north-west substructure and the main cluster. This suggests a merger between subclusters of similar masses taking place 
roughly in the plane of the sky.

\begin{figure*}
  \begin{center}
  {\includegraphics[angle=0,width=.9\textwidth]{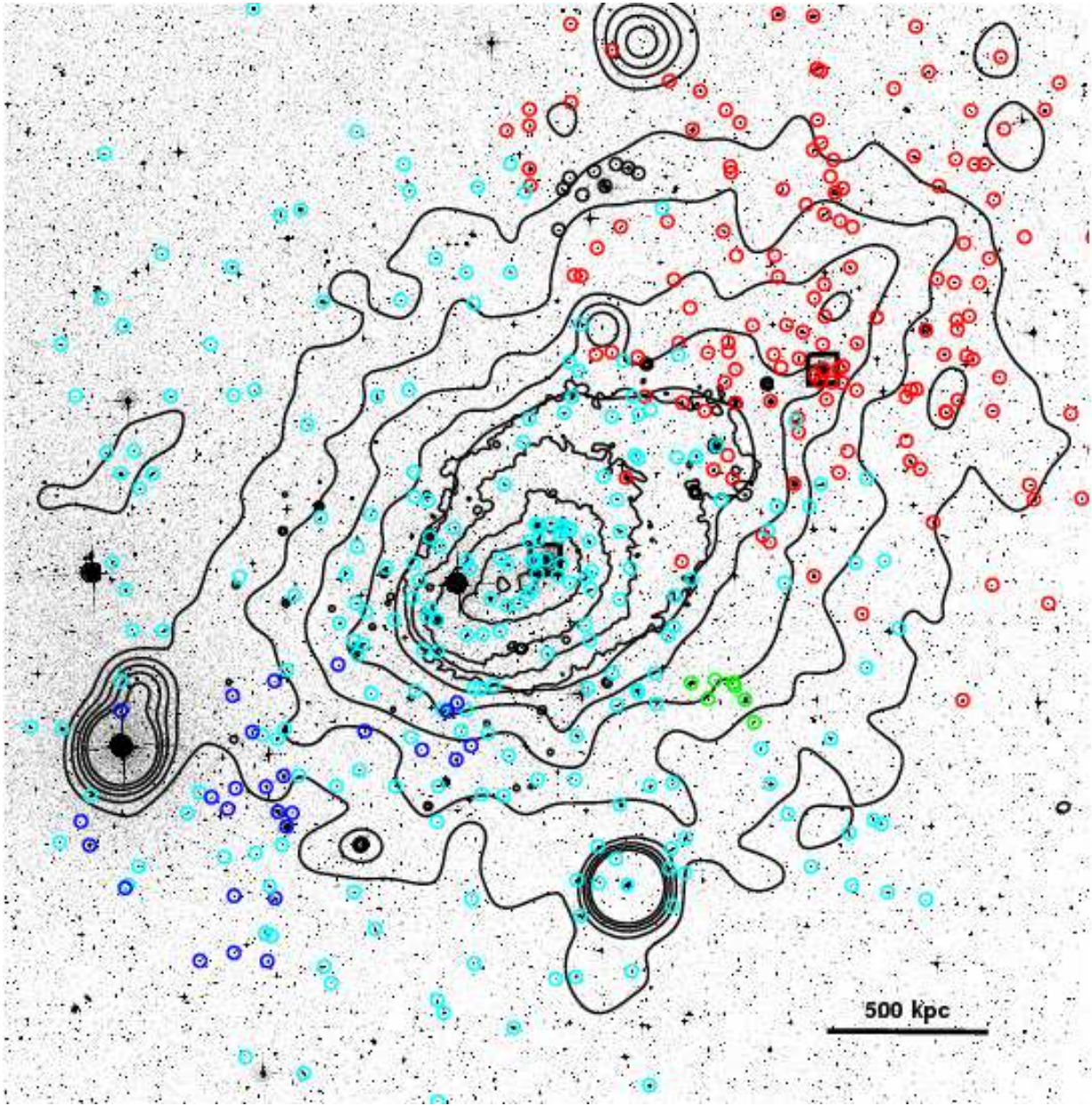}}
  \caption{Digital Sky Survey r-band image of Abell~3667. Circles mark the cluster members allocated to 
groups using the KMM method and are color coded to match the key seen in the {\it upper} panel of Figure~\ref{kmmalloc}. 
The black contours are from archival \chan\, (central 5 contour levels) and {\it ROSAT} (outer 5 contour levels) X-ray 
images. The cold front is visible as a compression of the X-ray contours to the south-east of the cluster center. The 
significant excess of emission associated with KMM2 noted by \citet{knopp1996} is visible as an elongation of the contours
toward the north-west, encompassing the KMM2 substructure. Black squares mark the positions of the two domninant cluster 
galaxies.}
  \label{dssimage}
  \end{center}
\end{figure*}

The direction of motion of the north-west substructure cannot be determined from the optical data alone, however it can 
be constrained by considering the radio and X-ray data in the literature. Regarding the radio data, the pertinent 
observation is the north-west radio relic described in \citet{rottgering1997} where inference can be made regarding the 
direction of motion by consideration of the spectral index and brightness distribution of the relic. 
\citet{rottgering1997} find that the relic brightness steepens at the north-west edge and is more diffuse to the 
south-east, while the spectral index is flat at the north-west edge and steepens towards the south-east. The model of 
\citet{roettiger1999} explains these features as due to a shock front moving from the south-east to north-west, where the 
steepening of the brightness and the flatness of the spectral index at the north-west edge occur because this is where the
shock acceleration of electrons is currently occurring. Since the shock moves from south-east to north-west, the electrons
to the south-east no longer have a source of energy for acceleration, and the spectral index steepening is caused by the 
aging of the particles as the shock moves off to the north-west. It should be noted, however, that recent observations 
\citep{mjh_thesis} have cast doubt on the existence of a flat spectral index at the north-west edge of the relic and
show only a moderate spectral gradient across the source. Even without these arguments, the curvature of the radio 
relic is consistent with a shock propagating to the northwest. The {\it ROSAT} X-ray observations of \citet{knopp1996} 
find significant evidence for extended X-ray emission associated with the north-west subcluster. In Figure~\ref{dssimage} 
we present the Digital Sky Survey r-band image of Abell~3667 with X-ray contours from \chan\, and {\it ROSAT} overlaid. The
contours clearly extend to encompass the north-west substructure. The fact that the subcluster 
hosts an X-ray halo, together with the arguments based on the radio data, suggest that the north-west subcluster has 
significant mass and is moving from south-east to north-west with the majority of its motion in the plane of the sky. 

In Figure~\ref{dssimage} the cold front is seen as a compression of the X-ray contours to the south-east of the cluster 
center. \cite{vikhlinin2001b} argued that the cold front is moving in a south-easterly direction at approximately the 
sound speed of the ICM. Based on arguments about the constant shape and the stability of the front, \citet{vikhlinin2002} 
predict the existence of a dark halo associated with the front, with its centroid $\sim 90$\,kpc from the leading edge of 
the front. We find no evidence for any structure in the galaxy distribution in this region, and note that the weak lensing
maps of \citet{joffre2000} do not show any significant mass concentrations here either. In the two-body merger scenario, 
the cold front is produced when core gas is `sloshed' out of the central potential well after the core passage of KMM2, 
and an excellent example of this is seen in \citet{poole2006}'s Figure~18 which shows entropy maps from the 3:1 mass 
ratio, offset impact parameter simulations. Interestingly, the structure that \citet{mazzotta2002} interpret as a 
hydrodynamic instability is seen in these simulations and interpreted as a `plume' of cool gas ejected from the main 
cluster's cool core due to the subcluster's pericentric passage. These simulations also produce two shock fronts on 
opposite sides of the main cluster moving in opposite directions, which would account for the observed radio relics in 
Abell~3667. This scenario is essentially the same as that proposed in \citet{roettiger1999}, although the simulations of 
\citet{poole2006} are at higher resolution and thus provide a better view of the cold front. In this scenario, KMM4 can be
interpreted in two ways; In the first, KMM4 is a group either infalling from the foreground, or is a background structure 
that is not physically associated with the cluster, nor with any of the X-ray and radio features or the merger. The second
interpretation is that KMM4 was once associated with KMM2, and was tidally stripped from the main KMM2 structure during 
its approach to the main cluster core and survives as a dynamically distinct substructure within the cluster.

An alternative scenario for the features seen in Abell~3667 involves a three body merger, between KMM2, KMM4 and KMM5 
where the arguments regarding KMM4 and its association with the X-ray and radio structures observed in the north-west 
presented above are essentially unchanged, however the cold front is the remnant core of KMM4, the centroid of which lies 
$\sim 550$\,kpc to the south-east of the front. A shock front driven by KMM4 is responsible for the radio relic in the 
south-east. In this scenario KMM4 has traveled from the north-west and passed through the cluster core where its ICM was 
significantly slowed by ram pressure, while the collisionless galaxies and dark matter continued their passage towards 
the south-east, similar to that seen in the Bullet cluster \citep{markevitch2002}. The mass maps of \citet{joffre2000} 
indicate a significant mass concentration not quite coincident with KMM4, but offset by $\sim300$\,kpc to the south-west. 
This lies well within the $2\sigma$ boundary seen in Figure~\ref{kmmalloc}, and the mass is highly likely to be 
associated with KMM4. The velocity dispersion of KMM4 is low and is unlikely to reflect the initial mass of the system, 
since the group must have been strongly affected by tidal forces and dynamical friction during its passage through the 
cluster core. The mass of the group was probably not comparable to the Abell~3667 system, since its gravitational 
potential well was unable to stop the gas from being completely stripped from the group \citep[see][for a comparison of 
the effects of different mass ratios on ram pressure stripping of core gas]{ascasibar2006}. 

\subsection{Summary and Conclusions}

We have presented an optical analysis of the cold front cluster Abell~3667 based on a large sample of (550) 
spectroscopically confirmed cluster members using data obtained on the AAT AAOmega facility. We find the cluster has a 
velocity dispersion of $\sigma=1056$\kms\, and redshift $z=0.0553$. Although the velocity distribution does 
not differ significantly from a Gaussian, the cluster shows clear evidence of subclustering in the galaxy density 
plots, and also contains significant local velocity substructure according to the k-statistic. Using the KMM algorithm, 
the cluster can be partitioned into 3 main components - the main cluster (KMM5) with 242 members, $\sigma=1073$\kms, 
$\overline v_{pec} = -103$\kms, a north-west subcluster (KMM2) with 164 members, $\sigma=1039$\kms, 
$\overline v_{pec} = 422$\kms, and a south-east subgroup (KMM4) with 27 members, $\sigma=209$\kms\ and 
$\overline v_{pec} = -638$\kms.

We confirm that Abell~3667 is a merging cluster, and the physical phenomena associated with observations at 
X-ray and radio wavelengths can be associated with substructure observed in the galaxy distribution, although it is 
apparent that detailed, combined N-body/hydrodynamical simulations are required to determine the plausibility of the two 
scenarios presented. This study highlights the need for large, deep spectroscopic samples to be used in conjunction with 
the full multi-wavelength observations when determining the dynamical state of a cluster of galaxies. Future work will 
involve using the spectroscopic sample to study the galaxy properties within Abell~3667, and to correlate them with 
substructures within the cluster.

We have now studied in detail two clusters from our sample of `cold front' clusters selected from the \chan\, archive,
finding here, as for Abell~1201 in Paper I, that despite difficulties in determining precise merger histories, cold 
fronts can be directly related to cluster merger activity and thus are reliable signposts of ongoing mergers. However, it 
must be noted that both Abell~1201 and Abell~3667 harbor signs of merger activity (other than cold fronts) in their \chan\,
images, thus were expected to show evidences of merger activity at other wavelengths. The acid test for this 
claim will come from detailed observations of cold front clusters with relaxed X-ray morphologies.

\section{Acknowledgments}

We thank David Woods and Gregory Poole for useful conversations. We also thank the staff at the Anglo-Australian 
Observatory for their support during the observations, and in particular Matthew Colless for offering Director's Time in 
order for this project to be undertaken. MSO acknowledges the support of an Australian Postgraduate Award. We acknowledge 
the financial support of the Australian Research Council (via its Discovery Project Scheme) throughout the course of this 
work. PEJN was supported by NASA grant NAS8-01130.

This research has made use of the NASA/IPAC Extragalactic Database (NED) which is operated by the Jet Propulsion 
Laboratory, California Institute of Technology, under contract with the National Aeronautics and Space Administration.

\label{lastpage}
\end{document}